%
%

\input harvmac.tex


\def\IR{\relax{\rm I\kern-.18em R}}
\def\IZ{\relax\ifmmode\hbox{Z\kern-.4em Z}\else{Z\kern-.4em Z}\fi}
\def\cD{{\cal D}}

%
%

%
%


\lref\bfss{T. Banks, W. Fischler, L. Susskind and S. Shenker,
``M Theory As A Matrix Model: A Conjecture'',
hep-th/9610043, Phys.Rev. D55 (1997) 5112-5128.}

\lref\tomrev{T. Banks, ``Tasi Lectures on Matrix Theory'',
hep-th/9911068.}

\lref\old{A. Polyakov, ``The Wall of the Cave'', hep-th/9809057,
 Int.J.Mod.Phys. A14 (1999) 645-658.}

\lref\juan{J.M. Maldacena, ``The Large N Limit of Superconformal Field 
Theories and Supergravity'', Adv. Theor. Math. Phys. 2, 231, 1998.}

\lref\otheb{E. Witten, ``Anti De Sitter Space And Holography'',
hep-th/9802150, Adv. Theor. Math. Phys. 2 (1998) 253-291.}

\lref\othea{S.S. Gubser, I.R. Klebanov, A.M. Polyakov, 
``Gauge Theory Correlators from Non-Critical String Theory'', 
hep-th/9802109, Phys.Lett. B428 (1998) 105-114.}

\lref\revi{O. Aharony, S.S. Gubser, J. Maldacena, H. Ooguri and Y. Oz, 
``Large N Field Theories, String Theory and Gravity'', hep-th/9905111.} 

\lref\bv{M. Berkooz and H. Verlinde, 
``Matrix Theory, AdS/CFT and Higgs-Coulomb Equivalence'', 
hep-th/9907100, To be published in JHEP.}

\lref\ofm{O. Aharony and M. Berkooz, ``IR Dynamics of d=2, N=(4,4) 
Gauge Theories and DLCQ of Little String Theories'',
hep-th/9909101, JHEP 9910 (1999) 030.}

\lref\natisix{N. Seiberg, ``Notes on Theories with 16 Supercharges'',
hep-th/9705117, Nucl. Phys. Proc. Suppl. 67 (1998) 158-171.}

\lref\vafaraj{C. Vafa, Talk given at IAS, spring 98; R. Gopakumar, 
Talk given at Aspen, summer 99.}

\lref\lennbh{L. Susskind and J. Uglum, ``String Physics and Black Holes'',
hep-th/9511227, Nucl. Phys. Proc. Suppl. 45BC (1996) 115-134 and
references therein.}

\lref\bd{M. Berkooz and M.R. Douglas, ``Five-branes in M(atrix) Theory'',
hep-th/9610236, Phys.Lett. B395 (1997) 196-202.}

\lref\abkss{O. Aharony, M. Berkooz, S. Kachru, N. Seiberg and E. Silverstein, 
``Matrix Description of Interacting Theories in Six Dimensions'',
hep-th/9707079, Adv. Theor. Math. Phys. 1:148-157,1998.}

\lref\edhgs{E. Witten, "On the Conformal Field Theory of the Higgs Branch",
hep-th/9707093, JHEP 9707 (1997) 003.}

\lref\abks{O. Aharony, M. Berkooz, S. Kachru and E. Silverstein,
``Matrix Description of (1,0) Theories in Six Dimensions'',
hep-th/9709118, Phys.Lett. B420 (1998) 55-63.}

\lref\sav{S. Sethi, ``The Matrix Formulation of Type IIB Five-branes'', 
hep-th/9710005, Nucl. Phys. B523:158-170,1998.}

\lref\gs{O.J. Ganor and S. Sethi ``New perspectives on Yang-Mills 
Theories with Sixteen Supercharges'', hep-th/9712071, JHEP 9801:007,1998.}

\lref\imsy{N. Itzhaki, J.M. Maldacena, J. Sonnenschein and S. Yankielowicz,
``Supergravity and Large N Limit of Theories with Sixteen Supercharges'',
hep-th/9802042, Phys. Rev. D58, 1998.}

\lref\natisix{N. Seiberg, ``Notes on Theories with 16 Supercharges'',
hep-th/9705117, Nucl. Phys. Proc. Suppl.67:158-171,1998.}

\lref\ks{A. Kapustin and S. Sethi, ``The Higgs Branch of Impurity Theories'',
hep-th/9804027, Adv. Theor. math. Phys. 2:571-591,1998.}

\lref\lngstr{L. Motl, 
``Proposals on nonperturbative superstring interactions'', hep-th/9701025;
T. Banks and N. Seiberg, ``Strings from Matrices'', 
hep-th/9702187,  Nucl.Phys. B497 (1997) 41-55.}

\lref\dvvvrtx{R. Dijkgraaf, E. Verlinde and H. Verlinde, 
``Matrix String Theory'', hep-th/9703030, Nucl.Phys. B500 (1997) 43-61.}

\lref\incre{W. Fischler, E. Halyo, A. Rajaraman and L. Susskind,
"The incredible shrinking torus",
hep-th/9703102, Nucl.Phys. B501 (1997) 409-426.}


%

\Title{\vbox{\baselineskip12pt  \hbox{hep-th/9912241}
				\hbox{PUPT-1907}}}
{\vbox{ \centerline{From SYM Perturbation Theory to Closed Strings}
        \centerline{}
	\centerline{in Matrix Theory} }}


\medskip

\centerline{\it Micha Berkooz
\footnote{${}^1$}{mberkooz@feynman.princeton.edu}
}
\bigskip

\centerline{Department of Physics, Princeton University, Princeton, NJ 08544}

\smallskip

\vglue .3cm
\bigskip

\centerline{Abstract}

\noindent For the purpose of better understanding the AdS/CFT 
correspondence it is useful to have a description of the theory for
all values of the 't Hooft coupling, and for all $N$. We discuss such
a description in the framework of Matrix theory for SYM on D4-branes,
which is given in terms of quantum mechanics on the moduli space of
solutions of the Nahm equations. This description reduces to both SYM
perturbation theory and to closed string perturbation theory, each in
its appropriate regime of validity, suggesting a way of directly
relating the variables in the two descriptions. For example, it shows
explicitly how holes in the world-sheets of the 't Hooft expansion
close to give closed surfaces.

\Date{12/99}

\newsec{Introduction}

In this note we discuss a Matrix model \bfss\ (for a recent review see
\tomrev) for the super-Yang-Mills
theory on D4-branes, and show how it transforms the SYM weak coupling
Hilbert space and interactions (in DLCQ) into a weakly coupled string
theory when N and the 't Hooft coupling become large.

Understanding the relation between gauge theories and strings has been
an interesting field of research for some time now (\old, and
references therein).  A breakthrough in this field is the AdS/CFT
correspondence \juan, in which the large N limit of certain gauge
theories (or otherwise interacting field theories) was solved in terms
of a gravity background.  Using this correspondence one can
study some properties of theories with gravity and of gauge theories
\othea\otheb\ (for a recent review see \revi).

This conjecture has by now been put to the test and verified in
numerous ways. Still, one might be interested in a more
straightforward way of deriving strings from SYM. Ideally one would
like to start perhaps with SYM perturbation theory (i.e., weak
coupled SYM where we understanding what are the ``fundamental'' fields
and their interactions) and re-sum the perturbation theory to strong
't Hooft coupling, or begin with Wilson loops, which are again best
understood at weak coupling, and again try and extend those to large
't Hooft coupling.

In this paper we will discuss a construction along these lines for the
case of SYM in 4+1 dimensions (with maximal supersymmetry). We will be
working at some large but fixed N. The description that we will use is
the Matrix description of this theory, which is valid for all values
of the 't Hooft coupling and $N$.  This is a quantum mechanical
system, which has the interesting property that in the form that we
will use it, it can be written in terms of both "closed string" and
"SYM" variables, on almost equal footing. Both sets of variables,
however, cannot be taken to be dynamical and independent at the same
time, as there are constraints which relate the two sets. One can
choose only one of these sets of variables to describe the system,
giving us either an "open string" (SYM) or an equivalent "closed
string" description of the same system. As expected, the "SYM"
variables describe the system better at weak 't Hooft coupling where
they give the SYM perturbation theory, and the "closed string"
variables describe the system at strong 't Hooft coupling and large N
where they give weakly coupled closed string perturbation theory on
the near horizon geometry. Again, these are two effective descriptions
of the same quantum mechanical system.

Since the AdS/CFT correspondence is a strong/weak coupling duality we
do not expect to have a simple map between the Hilbert space of weakly
coupled SYM (in DLCQ) and the Hilbert space of the closed strings (in
DLCQ). What we will obtain, however, is the best that one can hope for
- both SYM and closed strings perturbation theories in DLCQ are shown
to be the asymptotic expansions (in different regimes) of the same
underlying quantum mechanical system, of which we have a full
definition. Hence the duality between them is manifest in this
approach.
 
The organization of the paper is the following. In section 2 we
present the DLCQ (Matrix) model for D4-branes. In section 3 we analyze
it at weak 't Hooft coupling and briefly explain how SYM perturbation
theory is generated in this limit. In Section 4 we discuss the model
at strong 't Hooft coupling and large N - we review the near horizon
background of a cluster of D4-branes, and then obtain it by a
collective coordinate method in the DLCQ description. The method is
similar to that used in
\bv\ofm.
  
Before we proceed one should mention one caveat. SYM with 16
Supercharges in 4+1 dimensions exists\foot{Within field theories.  One
may also define it using "little string theories".} only as the 6D
(2,0) SCFT compactified on a circle (the size of the circle then sets
the SYM coupling) - for a review see \natisix.  This means that it is
not always easy to distinguish gauge theory large 't Hooft coupling
effects (i.e, when the loop corrections become strong at energies
$1/g^2_{ym}N$) from effects associate with the strong coupling of the
(2,0) field theory (which happens at some energy below
$g_{ym}^2$). This, however, will not play a role in what follows.

This work is reminiscent of ideas presented at \vafaraj. It is also
related to ideas in \lennbh, in which closed strings fragment to form
a gas of open string bits at the horizon (as in D-branes). The
construction presented here is a realization of this idea. We will see
the fragmentation quite explicitly, and how open and closed strings
transform into each other.

\newsec{The Matrix model of D4-branes}

Matrix theory can be used to describe the DLCQ of field theories on
certain solitons\foot{Typically, high dimensional high supersymmetry
ones.}. Initially one obtains a Matrix model for the theory on the
soliton coupled to gravity \bd. Then, in order to obtain a description
of the decoupled theory on the soliton, one traces the space-time
decoupling procedure in the Matrix model. The result is a
description of the soliton field theory
\refs{\abkss,\edhgs,\abks,\sav,\gs} in terms of a decoupled sector 
in the Matrix description. As usual in Matrix theory one then needs to
identify states with energy that scales like $1/P_{null}$ in the large
$P_{null}$ limit (where $P_{null}$ is momentum along the compactified
null circle), and only these can be compared with the states of the
uncompactified theory.  We would like to carry out this procedure for
$N_4$ D4-branes at fixed SYM coupling. By that we mean a cluster
of $N_4$ M5-branes compactified on a fixed circle, whose size is then
the 4+1D SYM coupling. In this section we will present the model, and
postpone its analysis in weak and strong 't Hooft coupling
to the next two sections.

The Matrix model for M5-branes in flat 11D space is
given in terms of the D0-D4 system \bd, and that of the compactified
M5-branes is given by essentially T-dualizing this system.  After
T-duality one obtains a system of $N_1$ wrapped D1-branes intersecting
(transversally) $N_4$ D3-branes with hypermultiplets living at the
intersection points. The D1-branes wrap a circle in $X^4$ direction
and the D3-branes span coordinates $X^0..X^3$, and may be located at
different points along the $X^4$ circle. More precisely, the Matrix
model is only the field theory on the D1-branes together with the
hypermultiplets. On a generic point of the circle the theory is the
standard theory of D1-brane, i.e., it has 16 supercharges. The
hypermultiplets live at fixed points on the circle and propagate in
time, i.e., they are impurities of codimension 1, and break half of
the supersymmetry. A detailed derivation, which we will not present
here, is found at \ks\ (and for the case of the Matrix model for 4D
${\cal N}=4$ SYM in \gs).  The system appears to be renormalizeable by
naive power counting arguments, but beyond that there is no good
understanding of its quantum mechanical properties. We will assume
that it is a consistent quantum theory.

The bosonic matter content is the following. The bulk (away from the
impurities) contains:

\noindent 1. A $U(N_1)$ gauge field, 

\noindent 2. 3 scalar field $X^i$ in the adjoint of $U(N_1)$. 
In the language of the D1-branes 
intersecting the D3-branes, these parameterize the positions of the
D1-branes in the spatial directions of the D3-brane. In the Matrix
interpretation, these 3 fields parameterize the positions in the 3
non-compact directions of the D4-brane, i.e., the coordinates of the
D4-branes other then time and null circle.

\noindent 3. 5 scalars $Y^\mu$ in the adjoint of $U(N_1)$. These 
parameterize the coordinates
transverse to the D3-branes (or D4-branes in the Matrix interpretation
of the model).

\noindent These fields are part of a single multiplet from the point 
of view of the 16 supercharges of the bulk. The full theory, however,
has 8 supercharges and the division of the bulk multiplet into the
${\cal N}=8$ multiplets is the following. Recall that in the D0-D4
system the coordinates of the D0 branes parallel to the D4 brane were in an
adjoint hypermultiplet. When we T-dualize parallel to the D4-brane,
these 4 fields generate the 3 scalar fields $X^i$ and the component of
the gauge field $A_1$, so all of these belong to the same ${\cal N}=8$
multiplet. The remaining fields are the analogue of a vector
multiplet as indeed they started their life as a 0+1 vector multiplet
in the D0-D4 system. This structure will manifest itself in the
coupling of the various D-terms below.

The impurities are hypermultiplets of the 8 supercharges, which
transform as fundamentals of $U(N_1)$. To specify the impurity data,
one specifies some points along the circle which the D1-brane wrap and
at each point the number, $n_i$, of hypermultiplets. These positions
correspond to the positions of the D3-branes along the $X^4$
circle. Each impurity point then has a global symmetry $U(n_i)$. The
total number of D4-branes is $\Sigma n_i$, and the different positions
of the impurities correspond to turning on null Wilson lines in the
D4-brane gauge group
\foot{Recall that the brane gauge theory appears as a global symmetry
in Matrix theory} which breaks it from $U(\Sigma n_i=N_4)$ to $\Pi
U(n_i)$.  We will actually restrict ourselves somewhat by turning on
null Wilson lines which break the gauge group to a product of
$U(1)$'s, i.e., each hypermultiplet impurity is located at a different
point. We will refer to this configuration as the ``resolved
model''. In the large null-momentum limit one expects that these
Wilson lines will not play a significant role (although it would be
interesting to better understand directly the non-resolved model).

The action that one obtains is the following. The bulk interaction
is:
\eqn\blkact{ \int dtd\sigma \biggl( \cD_t X^i\cD_t X^i + [Y^\mu,X^i]^2 + 
{1\over g_{ym}^2} \bigl(F^2+ \cD Y^\mu \cD Y^\mu +
[Y^\mu,Y^\nu]^2+{D^i}^2\bigr)+}
$$+{1\over g_{ym}} D^i\bigl(\cD_\sigma X^i+g_{ym} \epsilon_{ijk}
[X^j,X^k])\bigr)\biggr)+\ fermion\ terms$$ where $i=1,2,3$,
$\mu,\nu=1..5$ and $\cD$ without a subscript denotes summing over both
$t$ and $\sigma$. The impurities are located at points $\sigma_k,\
k=1..N_4$, and their interaction is
\eqn\impinta{\Sigma _k \int dt \biggl( (\cD_t Q^\alpha_k) 
(\cD_t Q^\alpha_k)^* + 
{Y(\sigma_k)}^2Q_kQ_k^* + {D^i(\sigma_k)} Q^\alpha_k \sigma^i_{\alpha\beta}
{Q^\beta_k}^* \biggr)+\ fermion\ terms} where $k$ is a $\Pi U(n_i)$
index (and we were not careful in lower and upper indices in cases
where it is clear how to raise and lower them).

The parameters of the 1+1 theory are related to those of spacetime in
the following way. Denoting by $\Sigma$ the size of the circle of the
1+1 theory, then
\eqn\param{\Sigma={1\over RM_s^2},\ g_{ym}^2={R^2M_s^4\over g_s^2}=
{R^2M_s^2\over g_{D4}^4} } where $g_{D4}$ is the SYM coupling on the
D4-brane.

In Matrix theory the coordinates $X,Y$ and $Q$ all measure spacetime
distances. This is obvious for the first two, but is also true for $Q$
(because $\sqrt{Q^2}$ is related to a certain instanton size). We will
denote the distance coordinates by a subscript $-1$, and then relation
to the fields in \blkact+\impinta\ is
$$X_{-1}=\sqrt{R\Sigma}X,\ Q_{-1}=\sqrt{R}Q,\ Y_{-1}= {
\sqrt{R\Sigma}\over g_{ym}} Y$$

\newsec{Some Comments on the Theory at Weak 't Hooft Coupling}

We are interested in discussing the dynamics of the Matrix model in
the limit of weak 't Hooft coupling on the D4-brane. Since this
coupling is dimensionful we actually require that the effective
coupling at the string scale be weak. This translates according to 
\param\ to the limit
\eqn\weakthoft{g_{ym}\rightarrow \infty} (keeping the energy fixed). 
In this limit we expect the dynamics to be governed by the moduli
space and configurations close to it. In section {\cal 3.1} we will
identify some configurations which remain in low energies in the open
string sector, and in section {\cal 3.2} we will discuss how
closed+open string perturbation theory comes about. In this section we
will be working at a large but fixed $N_4$.

\subsec{The long strings}

Given a Matrix model, the states which can be compared to those in the
uncompactified theory are those whose energy scales as $1/N_1$ in the
large $N_1$ limit. In this subsection we will discuss such states in
the regime when open string perturbation theory is valid, i.e., at
weak 't Hooft coupling.  Among these are of course the D4-brane
$U(N_4)$ gauge bosons, and although we will not focus on them
specifically in this section, it is clear how to do so.

Let us briefly remind the reader how weakly coupled string theory is
derived in the case of the flat 10D IIA string with no additional
solitons embedded in it \lngstr\dvvvrtx. To find low lying states one
begins by identifying appropriate pieces of the moduli space. One then
needs to identify small fluctuations around these configurations whose
energy has the correct $N_1$ scaling. Only these can then be compared
to the spectrum of the uncompactified theory. In the cases of impurity
systems the moduli space is rather complicated and is not understood
well enough for our purposes, but one can identify sometimes
excitations whose energy scales like $1/N_1$, without a need for a
detailed understanding of moduli space.

It is instructive to begin by identifying the states of the single
short open string, i.e., the states in the $N_1=1$ sector. In this
case one can actually identify the entire moduli space. The theory on
the D1 is a $U(1)$ gauge theory with 16 supercharges and in addition
there are charge 1 hypermultiplets localized at impurities (which
break half of the supersymmetries). The Higgs branch is the
following. In this case the D-term constraints are
\eqn\uonevac{ \partial_\sigma X^i(\sigma)=0,\ \sigma\not= \sigma_k}
$$ X^i(\sigma_k+)-X^i(\sigma_k-)=
g_{ym}Q^\alpha_k\sigma^i_{\alpha\beta}{Q^\beta_k}^*,\ \ k=1..N_4$$
where by $X(\sigma_k\pm)$ one means the limiting value of $X$ as one
approaches the point $\sigma_k$ from above (+) or from below (-). An
excited l-th hypermultiplet means that there are two strings ending on
the l-th D4-brane, with opposite charges with respect to the $U(1)$
which lives on that brane, and that the distance between their end
points on this brane is given by the 2nd line in \uonevac. Going to
the Higgs branch, i.e, turning on Q, also makes the $Y$ fields
massive.  The coordinates of the Higgs branch which we have identified
so far give us a space which corresponds to $N_4$ gauge bosons with
total charge $0$ in each $U(1)$, which are positioned at different
points on $R^3$. Additional coordinates of the Higgs branch are
related to Wilson lines made out of the gauge field $A_1$. These do
not play a role in the weak D4-brane 't Hooft coupling limit, as we
will discuss later.

An additional branch is of course the Coulomb branch in which $X,Y\not
=0$ but $Q=0$. This branch describes short strings in the bulk of the
IIA string theory.

There should also be additional ``mixed branches'', i.e., branches in
which the $Q$'s at only part of the impurities are non-zero (and zero
at the rest). This will have the interpretation of less then $N_4$
gauge bosons starting and ending on only some of the branes. This
however appears as a submanifold in the Higgs branch and it is less
clear how to think of it as an additional branch. Fortunately, this
problem happens primarily at the $N_1=1$ case, and is alleviated in
the long string picture, where obtaining the correct spectrum is more
critical. We will briefly discuss this later.

Next we identify the configuration of a collection of $p$ long
strings, all of which begin and end on the same brane (the
generalization to when they begin and end on different branes is
straightforward). We will not identify the precise classical zero
energy (flat directions) configuration, but rather a configuration
close to it with energy $1/N_1$. We will also require that the
different open strings are far apart from each other, i.e., if we
denote the average position of these long strings by
$X^i_1..X^i_p$ then our analysis will be to first order in
$g_{ym}^2Q^2/(X^i_{p_1}-X^i_{p_j})$. Since we are primarily interested in the
scaling with $N_1$ we will also set for now the size of the circle of
the 1+1 field theory to $2\pi$.

The 0'th order configuration is a configuration of the form
\eqn\cnfgzro{X_{0}^\mu(\sigma)= \Biggl(
\matrix{ X_1^i I_{m_1\times m_1} & & & & \cr
         & X_2^i I_{m_2\times m_2} & & & \cr & & . & & \cr & & & . &
         \cr & & & & X_p^i I_{m_p\times m_p} } \Biggr),} 
where $m_i$ are kept to be a fixed fraction of $N_1$ as we take
$N_1\rightarrow\infty$. We would now like to turn on the
hypermultiplets. Since we are dealing with a single brane we will turn
on only a single fundamental hypermultiplet, which we will take to
live at $\sigma=\pi$. Using the remaining $\Pi U(m_i)$ symmetries we
can rotate $Q$ such that only its first component in each block is
turned on, and we will also take ${\tilde Q}$ to be of the same form
(this will give us a rich enough family of approximate solutions).

Once we have turned on $Q$, we can choose the matrices at both sides
of $\sigma=\pi$ to be
\eqn\newmatx{X^i(\pi-)=X^i_0,\ \ 
X^i(\pi+)=X^i_0+g_{ym}Q\sigma^\mu Q^\dagger.}  To lowest order in $Q$
and $\tilde Q$, most of the eigenvalues of $X^i(\pi+)$ are the same as
those of $X^i(\pi-)$ except that one eigenvalue in each block is
shifted by order $Q^2$. It is also clear that we can discuss, to this
order, each block separately. Let us focus on the first block. We
would like to construct long strings, i.e., complete the configuration
to $X^i(\sigma)$ for all $\sigma$ such that the energy is proportional
to $1/m_1$.  It is clear that the appropriate long string
configuration is such that it begins in, say, $X_1(\pi+)$, winds $m_1$
times around the circle and ends at $X_1(\pi-)$. The minimal gradients
are of order $g_{ym}QQ^\dagger/m_1$, and since the length of the long string
is $m_1$,the total energy scales like $1/m_1$ which is the correct
scaling to be interpreted as physical state in Matrix theory. This
state would be that of a string starting and ending on the same brane.

We can now also discuss the situation in which some impurities are
activated and some are not. In the $N_1=1$ case this was somewhat
problematic because we were looking for these states as wave functions
on an exact moduli space. In the large $N_1$ limit we relaxed this and
it is easy to understand where the additional required states come
from. In this case we are allowed to set $Q\not=0$ at some impurities
and $Q=0$ at the other if we set $Y=0$ at the first group and
$Y\not=0$ at the second. We can do so while keeping the gradients of
$Y$ small (i.e., scaling like $1/N_1$), such that they are
compatible with the requirement that the total energy scales as
$1/N_1$.

Since we have worked in the Matrix model of the full string theory 
then this sector contains both the gauge bosons, as well
as all the excited string states. However, it is clear that if 
we mimic the space-time decoupling limit in our model, then we
will end up only with the gauge bosons. 

One more complication is the following. As explained in \ks\ the
classical moduli space of solution to the D-term equations is actually
$4N_1N_4$ dimensional, whereas we have identified a space which can
have at most $3N_1N_4$ parameters which corresponds to all the
segments of the strings splitting (although for Matrix theory
applications we were interested in yet a smaller space). The remaining
$N_1N_4$ parameters are associated with the component $A_1$ of the
gauge field. This is to be expected based on what was explained before
that $A_1$ should actually be viewed as part of a hypermultiplet. A
convenient gauge invariant parameterization of these coordinates can
be given by a subset of the quantities
\eqn\cmpctdir{Q_i exp^{\int A_1 d\sigma}X^{i_1} 
exp^{\int A_1 d\sigma}X^{i_2}... Q^\dagger_{i+1}} where the $i$ and
$i+1$ index means that we compute the Wilson line between two
neighboring impurities. However for fixed $X^i$ this coordinate is
compact, and its kinetic term is multiplied by $1/g_{ym}^2$. Hence the
non-homogeneous wave functions along this direction will have an
energy proportional to $g_{ym}^2$ which in spacetime means a mass
proportional to $1/g_s$, i.e, they are not perturbative string
states. This is familiar from the study of the DLCQ closed string
field theory where exciting the Wilson loop in the 16-supercharges 1+1
SYM corresponds \incre\ to D-objects.

We have briefly noted before how the Dirichlet boundary conditions
come about (i.e., $Q\not=0$ requires $Y=0$ on or close to the moduli
space). The Neumann boundary conditions come about in the following
way. Let us focus on the case of $N_1=1$ (which is anyhow similar to
what we obtain after we go to the long strings). Using the 2nd line of
\uonevac\ we can now solve for $Q$ in terms of $X$ and insert it into
the Lagrangian (more precisely, we can solve for $Q$ up to a phase
which is a gauge degree of freedom). The resulting Lagrangian (placing
the "active" impurity at $\sigma=0$) is
\eqn\sublag{ 	\int_{R_-} (\partial X^i)^2 dtd\sigma +
 		\int_{R_+} (\partial X^i)^2 dtd\sigma + }
$$ + {1\over
g_{ym}} \int f_{ij}(X)   (\partial_t X^i(0+,t)- \partial_t X^i(0-,t))
			 (\partial_t X^j(0+,t)- \partial_t X^j(0-,t))
dt,$$
 where $f$ is some function which satisfies
$$f(\lambda X)={1\over \lambda} f(X).$$ The variation of the action
with respect to the ``boundary terms'' gives the equation
 $\delta X(0+)$ and $\delta X(0-)$ is
$$\delta X(0+)(\partial_\sigma X)(0+) - \delta X(0-)(\partial_\sigma
X)(0-) +{1\over g_{ym}}(....)=0.$$ The details of the last term are
irrelevant except that it is non-singular for $X\not=0$, i.e., at
generic points along the flat directions. Hence in the limit
$g_{ym}\rightarrow \infty$ it disappears. It is clear that the path
integral contains all values of $X(0\pm)$ and that the different
values of $X(0\pm)$ can be connected by physical processes (because
this is the case on the moduli space, as measured by its metric). This
implies that we can't consistently set $\delta X(0\pm)=0$. Hence the
effective boundary condition that we obtain when the string splits
along the impurity is
\eqn\neum{\partial_\sigma X^i(0\pm)=0,}
which is the correct Neumann boundary condition.

We have therefore shown how the states in the open string theory, 
and in particular those of the gauge multiplets in the D4-brane, come 
about in the Matrix model. And we have studied how their relative 
position is associated to the value of the impurity variables. Next,
we would like to outline how string and SYM perturbation theories, in 
weak 't Hooft coupling, come about (and what happens to them at large 
't Hooft coupling).

\subsec{Weak 't Hooft coupling perturbation theory}

String and SYM perturbation theories come about now in the following
way. The theory contains a large number of different branches, in
which different modes of the impurities are excited. Open+closed
string perturbation theory comes about as one passes from one branch
to another. As we saw above, an excitation of an impurity (i.e., going
to the branch $Q\not=0$) corresponds to opening a hole in the
world-sheet where Dirichlet boundary conditions are imposed on
coordinates transverse to the brane, and Neumann boundary conditions
on those parallel to the brane. If we are starting with some state,
then standard quantum mechanical perturbation theory tells us that we
have to sum over all ways of exciting an impurity, letting it decay,
exciting some other impurity etc. If we start with a closed string
state, this corresponds to summing over all ways of cutting holes in
the surface. In the finite $N_1$ system we are allowed to start these
holes on a discrete set of points on the surface. However, in the long
string picture there are $N_4$ such points within each interval of
size $1/N_1$ of the long string world-sheet. Hence in the large $N_1$
limit we can open a hole everywhere in the world-sheet, and summing
over all such opening (which is the same as summing over the different
ways of exciting the impurities) corresponds to summing over all the
moduli of the holes on the world-sheet.

Hence, the open string perturbation theory, i.e., the expansion in the
number of holes on the world-sheet is an expansion in the total number
of excited impurities. The transition from a branch in which an
impurity is not excited to a branch in which it is excited is mediated
by some operator, which should give the 't Hooft coupling $g_sN_4$
dependence of this transition (similar to \dvvvrtx). We have not
carried out this computation but the factor of $N_4$ is easy to
understand. In the resolved model it merely counts the number of
different impurities which can be excited, in an exact correspondence
to which brane is the end point of the string.  The factor of $g_s$ is
associated with exciting a single fixed impurity and therefore there
is no additional $N_4$ dependence.
 
Jumping ahead, let us now consider what happens when the 't Hooft
coupling becomes large. In this case the impurities are excited
frequently and an expansion in the number of impurities excited is no
longer useful. However, one should now think about the impurities as
new ``closed string'' degrees of freedom in the sense that as
$N_1\rightarrow\infty$ the impurities become dense (and evenly spread)
on the world-sheet of the long string. These are unusual "closed
string" degrees of freedom because, for example, there are no
$\partial_\sigma Q$ terms in the Lagrangian. Going to the
appropriate collective coordinates, we will see that their effective
dynamics is that of a closed string moving on the near horizon limit
of the D4-brane.

\newsec{The large 't Hooft coupling limit}
 
We would like to show how the theory of closed strings on the near
horizon limit of the D4-brane comes about from the Matrix model
described above, and how the dynamics of the impurity system becomes
the dynamics of closed strings. For this purpose, unlike the analysis
before, it is useful to first decouple gravity, and only then put the
remaining quantum mechanical system into the form of closed
strings. The derivation of the decoupled model is carried out in
section {\cal 4.2} and the derivation of the near horizon closed
string description is carried out in section {\cal 4.3}. But first we
would like to briefly review the near horizon limit of the D4-branes.

\subsec{The Near-Horizon background}

In this subsection we will briefly review this near-horizon background
of a cluster of $N_4$ D4-branes, following \imsy.

The near horizon limit appropriate for the D4-brane is:
\eqn\dlimt{Y^\mu_{nh}={r^\mu\over\alpha'}=fixed,\ 
\ g_{D4}^2=g_s\sqrt{\alpha'}=fixed,\ \ 
\alpha'\rightarrow 0,}
where $r^\mu$ are the coordinates transverse to the brane, $g_{D4}$ is
the Yang-Mills coupling on the D4-brane, and we neglected numerical
factors of order 1 (This limit may also be understood as that of an
M5-brane wrapped on a circle of size $g_{D4}^2$ which is held fixed as
$M_{p,11}\rightarrow \infty$ which gives the 6D $(2,0)$ CFT on a
circle). 

The corresponding type IIA background is (in string metric):
\eqn\twoabck{
ds^2=\alpha'\bigl( {Y_{nh}^{3\over2}\over g_{D4}\sqrt{N_4}}dx^2+
{g_{D4}\sqrt{N_4}\over Y_{nh}^{3\over2}}dY^2_{nh} + g_{D4}\sqrt{N_4Y_{nh}}
d\Omega^2\bigr)}
$$e^{\phi}=\bigl( {Y_{nh}^3g_{D4}^6\over N_4}\bigr)^{1\over4}.$$ The
type IIA solution can be trusted in the regime
\eqn\trustsol{N_4^{-1}<<g_{D4}^2Y_{nh}<<N_4^{1\over3}.}
For values of $Y_{nh}$ larger then the upper bound above, the coupling
is large and one needs to lift the solution to M-theory, where it
asymptotes to the near horizon limit of the M5-brane \juan\
compactified on a circle (which reflects the fact mentioned before
that the UV fixed point of the D4-brane is actually the 6D (2,0) fixed
point). For values of $Y_{nh}$ smaller than the lower bound in
\trustsol\ the curvatures become large, i.e, the world-sheet becomes
strongly coupled which reflects the fact that the 4+1 SYM becomes
weakly coupled in the IR. It would seem that at the lower end of the
region \trustsol\ the 4+1 theory is strongly coupled because the 4+1
effective 't Hooft coupling is large, rather then exhibiting a strong
coupling behavior associated with the (2,0) fixed point - the latter
takes over at the upper end of this region. Correspondingly the dual
description in this regime is given in terms of a string theory.

\subsec{The decoupled theory}

Equation \param\ describes the relation between the the parameters of
the Matrix model and the parameters of the type IIA string
theory. Hence it is straightforward to follow the decoupling limit
\dlimt\ in the Matrix model.  One more useful ingredient is a convenient 
scaling of the various fields in the 1+1 impurity Lagrangian in this
limit. The coordinates $X_{-1}$ and $Q_{-1}$ remain fixed, which
implies that we actually need to rescale $X$ (but keep $Q$ fixed). The
coordinate $Y$ (or $Y_{-1}=r$) is rescaled according to \dlimt\ such
that $Y_{nh}$ is kept fixed. To summarize we keep fixed the
coordinates
\eqn\fxdcoor{Q_{-1}=\sqrt{R}Q,\ X_{-1}=\sqrt{R\Sigma}X,\ 
Y_{nh}={1\over\sqrt{R\Sigma} g_{ym}} Y}
while taking the limit
\eqn\morelim{\Sigma\rightarrow 0,\ g_{ym}\rightarrow \infty,\ 
\Sigma g_{ym}^2 ={R\over g_{D4}^4}\ fixed,}
which, using \param\ amounts to $M_s\rightarrow\infty$ with $g_{D4}^2$
fixed.  In particular we see that $Y_{nh}$ differs from $Y$ by a
finite normalization which means that we basically keep $Y$ fixed in
this limit. The latter scaling is familiar from other cases in which
the near horizon Coulomb branch is identified with part of the Higgs
branch (in a fashion that is set by the hypermultiplet-vector
couplings) \bv\ofm.  Since the size of the circle is also rescaled, it
is convenient to choose a new coordinate $\sigma'$ which remains
finite, i.e.,
\eqn\coorscl{\sigma'={\sigma\over\Sigma}.}
Finally we rescale the gauge fields similarly to the coordinates, 
i.e., $A_t$ is not rescaled, and $A_{\sigma}$ is rescaled the
same way as $\partial_\sigma$.

Using the new quantities the bosonic part of the action becomes (and 
dropping the $(-1)$ subscript)
\eqn\nwacblk{ \int dtd\sigma' \biggl(
{1\over R} (\cD_t X^i)^2 + {R\over g_{D4}^4} [X^i,Y_{nh}^\mu]^2 +
R(\cD_{\sigma'} Y_{nh})^2+
  D^i\bigl( {g_{D4}^2\over R} (\cD_{\sigma'} X) + 
     {1\over R}\epsilon_{ijk}[X^j,X^k]^2\bigr) + }
$$ + {g_{D4}^2\over R} F^2 \biggr) +{\Sigma\over g_{ym}^2} \bigl( 
	{R^2\over g_{D4}^4} (\cD_t Y_{nn})^2+
	{R^4\over g_{D4}^4} [Y_{nh},Y_{nh}]^2 + {D^i}^2\bigr)+$$
$$+\Sigma_k\int dt {1\over R}\bigl( {(\cD_t Q_k)}^2 + 
D^i(\sigma_k)Q_k\sigma^i Q_k^* + 
{R\over g_{D4}^4} {Y_{nh}(\sigma_k)}^2Q^2\bigr)$$
where the sum is over the points of the impurities.

In the limit \morelim\ the kinetic terms for the vector multiplet $Y$
drops out, and it should be regarded as an auxiliary variable (in the
gauge $A_0=0$, $F^2$ is a kinetic term for a hypermultiplet field). If
we integrate it out we obtain the quantum mechanics of the Higgs
branch. To obtain the near horizon geometry, however, we follow the
procedure of integrating out the $Q$ hypermultiplets, and describe the
model in terms of an effective closed string theory \bv\ofm.

\subsec{The effective action for the vector multiplet}

We would like to integrate out the hypermultiplets $Q$ and obtain an
effective description for the $X$ and $Y$ fields. Since these are 1+1
fields, we will obtain a string theory, which of course will be the
type IIA DLCQ Matrix string field theory on the near-horizon geometry
of the D4-brane. Since for fixed $Y,X$ and $D$ the $Q$'s appear
quadratically, it is straightforward to do the integration. We will
expand\foot{At large values of $Y$, where the string coupling is
large, we can still go to the non-abelian form of the Matrix string
field theory on the near horizon limit. We will not, however, be able
to go reliably to the long strings picture.} the result in $\partial
Y/Y^2$.

It is convenient to begin with the case that the 1+1 field theory is a
$U(1)$ gauge theory and take the number of D4-branes to infinity. It
will then be clear how the $U(N_1)$ case works even for finite $N_4$,
which is actually our final goal. Again we will be working in the
regulated model, in which all the impurities are separated. For the
case of the $U(1)$ theory we will also assume that the impurities are
scattered more or less uniformly around the circle of the world-sheet.

\medskip

\noindent{{\it 4.2.1} The $U(1)$ effective action}

\medskip

For a $U(1)$ gauge theory one drops all the commutator terms from the
action \nwacblk. Integrating out the $Q$ variables is straightforward
and gives (dropping the decoupled $F^2$)
\eqn\uoneintga{\int dtd\sigma' \biggl( {1\over R}(\partial_t X)^2 + 
R(\partial_{\sigma'} Y_{nh})^2 + {g_{D4}^2\over R} D^i
\partial_{\sigma'} X^i
\biggr) +}
$$+\Sigma_k \int dt {g_{D4}^2\over R{Y_{nh}(\sigma_k)}^3} 
(\partial_t Y_{nh}(\sigma_k))^2 +
{g_{D4}^6\over R^3}{D(\sigma_k)}^2$$ \
The next step is to note that as the number
of impurities goes to infinity and their location becomes dense on the
circle, then we can replace the sum over impurities by an
integral. After integrating out the D-term we obtain a string
action, whose bosonic part is
\eqn\uonestrng{\int dtd\sigma' \biggl( 
{1\over R} (\partial_t X)^2 + {R\over g_{D4}^2N_4} Y_{nh}^3
(\partial_{\sigma'} X)^2+ {g_{D4}^2N_4\over R} Y_{nh}^{-3}(\partial_t
Y_{nh})^2 + R(\partial_{\sigma'} Y_{nh})^2\biggr)}

A string action of this form is somewhat less familiar since it is not
in the usual gauge
$\gamma^{\alpha\beta}=\delta^{\alpha\beta}$. However, it is easy to
read the world-sheet metric and the target space metric from this
action. The former is
\eqn\wrldmtrc{\sqrt{\gamma}\gamma^{\alpha\beta}= \biggl(
\matrix{ {g_{D4}N^{1\over2} \over Y^{3/2}} & 0\cr
	  0 & {Y^{3/2}\over g_{D4}N^{1\over2}} \cr}
\biggr) }
(where the 1st component is the time) and the latter is:
\eqn\wrldsht{ {Y^{3\over2}\over g_{D4}N^{1\over2}}  dx^2 + 
              { g_{D4}N^{1\over2} \over Y^{3\over 2}} dy^2} which is
the string metric of the D4 background.

It is worth explaining the choice of gauge for the world-sheet metric,
Especially since in the gauge \wrldmtrc\ it is set to a field (Y)
dependent value.  This is actually natural in the context of light
cone quantization in this type of background. Consider the action
before fixing reparametrization invariance and the light cone
condition. The relevant part for our purposes is
\eqn\pmact{\int dtd\sigma \sqrt{\gamma} \gamma^{\alpha\beta} 
\partial_\alpha X^+\partial_\beta X^- Y^{3\over2}.}
We would like to impose the gauge $X^+=\tau$. In order to do that we
need that $\tau$ will be a solution of the equations of motion for
$X^+$, i.e., $\partial_\alpha Y^{3\over2} \sqrt{\gamma}
\gamma^{\alpha\beta} \partial_\beta \tau=0$, which implies that we
need to set
$$\sqrt{\gamma}\gamma^{00}\propto Y^{-{3\over2}}.$$

\noindent{ {\it 4.2.2} The $U(N_1)$ case }

The Matrix prescription instructs us to take $N_1\rightarrow\infty$,
which is what we turn to now. We will then pass to the picture of the
long strings, which is again similar to the $U(1)$ case above. This
will also clarify why we mandated in the $U(1)$ case (a short string)
that the number of impurities goes to infinity and that they are
evenly spread around the circle. Whereas this was arbitrary in the
$U(1)$ case we will see that this is automatically true for the theory
that lives on the long string.
 
When discussing the $U(N_1)$ theory one needs to restore the
commutator terms and the covariant derivatives in
\nwacblk. Furthermore the terms $Y^3(\partial X)^2$ and
$Y^{-3}(\partial Y)^2$ will be replaced by non-abelian generalizations.
Another change is that we will also generate a term which is roughly
of the form
$$\propto \Sigma_k \int dt {1\over Y^3}[Y,Y]^2.$$ This commutator term
in addition to the term $[Y,X]^2$, which is one of the
commutators we initially had, tell us that the generic flat directions
are when all the $X$ and $Y$ matrices commute. Hence we can go to the
"long string" \lngstr\dvvvrtx, which is a $U(1)$ string with length
$N_1$.

When we go to the long string we are also instructed to concentrate on
very long wave length, i.e., on wave lengths of order $N_1$. Relative
to these wave lengths the impurities are dense since the separation
between them is of order 1. In the $N_1\rightarrow\infty$ we are
justified (even for a finite number of D4-branes) to go to the
``continuum impurities'' approximation as we did when we went from
\nwacblk\ to
\uonestrng, leaving us with a final result \uonestrng\ as the effective
dynamics of the long string.
 
Hence we obtained what we were looking for, i.e., within the impurity
model we were able to identify configurations of long strings which
are governed by a sigma model on the near horizon background of the
D4-brane.

As in \dvvvrtx,\ofm\ we can also estimate the behavior of the string
coupling. To do so we need to identify the mass scale set by the
coefficient of the commutator term. The inverse of this mass scale
will then determine the string coupling \dvvvrtx. To correctly
identify the mass scale we would like to rescale the coordinates such
that the world-sheet metric is (locally) the canonical metric, and then
rescale the fields such that their kinetic term is normalized.  We
will, arbitrarily, focus on the $X$ coordinates. The terms in the
Lagrangian that contain only $X$ coordinates are (neglecting $N_4$ and
$g_{D4}$ dependence)
\eqn\xlag{\int dtd\sigma' 
\biggl( (\partial_t X)^2 + Y^3(\partial_{\sigma'} X)^2
+Y^3[X,X]^2 \biggr) } To go to a Lorentz invariant form with
canonically normalized kinetic term we rescale
$$\sigma'=Y^{3\over2}\sigma'',\ \ X=Y^{-{3\over 4}}{\hat X}$$
to obtain the action
\eqn\xlagb{\int dtd\sigma'' \biggl( (\partial_t {\hat X})^2 + 
(\partial_{\sigma''} {\hat X})^2 +Y^{-{3\over2}}[{\hat X},{\hat X}]^2
\biggr).}  We can now read the string coupling from the
coefficient of the last term to be
\eqn\cpl{g_s(Y)\propto Y^{3\over4}}
which is the correct dependence in \twoabck.

To conclude, we have used Matrix theory to formulate SYM on D4-branes
in a way that contains both open and closed string variables. The
closed string variables are auxiliary variables and when we integrate
the out, we obtain the open string description and SYM perturbation
theory in weak 't Hooft coupling. When we choose to work with a
different set of variables, i.e, the closed string variables, we get
in a straightforward manner the description of closed strings on the
near horizon geometry (both metric and string coupling). The fact that
the two sets of variables are related in a fairly simple way, and that
the entire procedure is done in quantum mechanics suggests a simple
way of microscopically identifying states in the two descriptions.

\bigskip

\centerline{\bf Acknowledgments}

I would like to thank O. Aharony, A. Giveon, R. Gopakumar, D. Kutasov,
G. Lifschytz, O. Pelc, M. Rozali, N. Seiberg, S. Sethi,
E. Silverstein, E. Verlinde and H. Verlinde for useful discussions,
and particularly A. Kapustin for collaboration at early stages of this
work. I would also like to thank the Weizmann Institute of Science for
its hospitality during the completion of this work. This work is
supported by NSF grant 98-02484.

\listrefs

\end